# Photoluminescence study of V-groove quantum wires:
## The influence of disorder on the optical spectra and the carrier thermalization

D.Y. Oberli, F. Vouilloz, R. Ambigapathy, B. Deveaud and E. Kapon

Department of Physics, Swiss Federal Institute of Technology Lausanne

CH-1015 Lausanne-EPFL, Switzerland

Email : Daniel.Oberli@epfl.ch

Fax : 0041-21-693-4525

Tel. : 0041-21-693-5486

### ABSTRACT

We report on time-resolved and steady-state photoluminescence studies of GaAs/AlGaAs V-groove quantum wire structures. Steady-state photoluminescence experiments are performed in the temperature range from 8K to 200K. We evaluate the relation between photoluminescence excitation and absorption and determine experimentally an optical density in order to analyze the temperature dependence of the photoluminescence spectra. We find that, at a temperature above 60K, the photoexcited electron-hole pairs reach a thermal equilibrium at the lattice temperature while, at a temperature below about 60K, they do not reach a quasi-equilibrium in the steady-state. Time-resolved photoluminescence studies performed at a carrier density of about $2.10^4$ $cm^{-1}$ indicate that, at 60K, a quasi-equilibrium is reached on a time scale of 10 ps. Furthermore, the hot carriers cool in about 100 ps to the lattice temperature. At 8K, however, evidence of a non-thermal carrier distribution is found at the earliest times, which suggests that carriers in extended states are not in thermal equilibrium with carriers in localized states.

## 1 Introduction

Semiconductor quantum wires (QWRs) are unique systems to explore one-dimensional (1D) quantum effects on their electronic and optical properties. QWRs produced either by cleaved-edge overgrowth (T-shaped wires) or by growth on patterned substrates (V-groove wires) have attained high structural and optical quality. In most optical studies, however, the omnipresent structural disorder in high quality nanostructures leads to localization of excitons and modification of their radiative properties [1]. Evidence for localized excitons has been found in spatially-resolved photoluminescence studies [2,3,4].

Relaxation dynamics of carriers are strongly influenced by the reduced phase space available for carrier-carrier scattering in one dimensional systems. Owing to the constraints imposed by conservation of energy and momentum in a scattering event, electron-electron scattering is expected to be largely suppressed in quantum wires. The inclusion of non-Markovian electron-electron quantum kinetics was shown to enhance electron-electron scattering processes and lead to fast intrasubband thermalization at electron densities above about $10^4$ cm$^{-1}$ [5]. We note that the inhibition of intrasubband e-e scattering in the Born approximation disappears when non-parabolic terms in the dispersion are included; however, the scattering rate of these processes is negligibly small [6]. Furthermore, quantum wires in GaAs are all oriented along a $(1\bar{1}0)$ low symmetry axis and, thus, the conduction band is spin split due to the lack of inversion symmetry. Consequently, electron-electron scattering is expected to be restored across the two different spin subbands leading to new spin-related phenomena [7]. The influence of triple-electron intrasubband collisions on the relaxation of an 1D electron gas was evaluated theoretically [8]. It was shown to be an effective scattering process in a non-degenerate 1D electron gas at densities of about $10^5$ cm$^{-1}$. The relaxation of hot carriers into the lowest-energy subbands is also influenced by phonon intersubband emission as well as carrier-carrier intersubband scattering when electron-hole pairs are photoexcited on several subbands [9]. It can be expected, however, that disorder will influence the redistribution in k-space of photoexcited carriers by providing a phase breaking mechanism. In the present paper we focus on the thermalization of photoexcited carriers in a QWR and present experiments to unravel the relative importance of these scattering processes.

## 2 Experimental details

We studied a high-quality QWR sample grown by low-pressure organometallic chemical vapor deposition on a V-grooved (001)-GaAs substrate; more details on the growth and morphology of such a sample are given elsewhere[10]. The nominal GaAs layer thickness was 2.5 nm resulting in a thickness at the crescent center of 7 nm. The GaAs QWR was embedded in $Al_{0.3}Ga_{0.7}As$ barrier layers. The upper part of the sample was chemically etched away including the top and part of the side quantum wells in order to reduce strongly their contributions to the photoluminescence spectrum. The sample was mounted onto the cold finger of a continuous-flow helium cryostat. The steady- state photoluminescence experiments were performed over a temperature range from 10K to about 200 K by exciting the sample at a wavelength of 700 nm with a titanium-sapphire laser operating in a continuous wave (cw) mode. We used a constant and low power density of 25 $W/cm^2$ over the whole temperature range. In the time-resolved photoluminescence experiments it was excited with 2 ps light pulses from a mode-locked titanium-sapphire laser at a wavelength of 730 nm and a power density of 50 $W/cm^2$. The carriers were generated below the AlGaAs barriers at this wavelength. The luminescence was detected using a synchroscan streak camera with a temporal resolution set at 10 ps in order to increase the sensitivity of the experiment.

## 3 Experimental results and analysis

Photoluminescence spectra of the QWR sample at several different temperatures are displayed in Fig. 1 for cw excitation. The luminescence spectra of the GaAs QWR are characterized by a dominant peak arising from the excitonic optical transition ($e_1$-$h_1$) associated with electrons and holes in their respective 1D subband of lowest index. On the high energy side of that peak one finds contributions to the QWR emission from the higher optical transitions associated with higher index subbands. This high energy tail broadens and extends farther in energy from the main peak as the temperature is raised, as is expected from the recombination of a heated carrier distribution. We find that, above a temperature of about 100K, the main peak of the QWR luminescence starts to follow the known temperature dependence of the GaAs energy band gap. At a lower temperature, however, the PL peak is shifted down in energy with respect to the expected temperature dependence of the energy gap. At 8K, this energy shift amounts to 6.4 meV, a value that is nearly equal to the measured Stokes shift (5.4 meV) as previously reported for other QWR samples [11]. This energy shift can thus be attributed to disorder related effects. If the emission at high temperature had been dominated by the recombination of correlated electron-hole pairs above the band edge it would have resulted in an energy shift larger than the exciton binding energy at 8K (about 11 meV). We infer from this observation that the emission of the QWR is dominated by excitonic recombination over the whole temperature range since at low temperature the excitonic nature of the emission in cw-PL spectra at low density is well established. When examining the PL spectra and their extended high energy tails the question that arises is whether the photoexcited carriers have reached a thermal equilibrium with the lattice or alternatively a quasi-equilibrium among themselves.

We then consider the fundamental relationship between absorption and spontaneous emission rates assuming that the photoexcited carriers have reached a quasi-equilibrium described by a temperature T and a chemical potential μ [12]. This relation is essentially given by :

(1)   $I_{PL}(E) \propto \alpha(E) \cdot \dfrac{1}{\exp((E-\mu)/k_BT) - 1}$

where α is the absorption at the photon energy E and $I_{PL}$ the emission intensity. If the density of photoexcited carriers is low then relation (1) can be rewritten as :

(2)   $I_{PL}(E) \propto OD_T(E) \cdot \exp(-E/k_BT)$

where $OD_T(E)$ only differs from the absorption by a constant factor and the Bose distribution function has been replaced by a Boltzmann factor. It is well known that absorption spectra are in general not equivalent to photoluminescence excitation (PLE) spectra. Because a direct measurement of an absorption spectrum is difficult to perform on an array of V-groove QWRs (the GaAs substrate is opaque) we instead determine the experimental optical density from the measured PL spectrum at 100K by assuming that the photoexcited carriers have reached a thermal equilibrium at the substrate temperature. The optical density at 100K is depicted in Fig. 2 together with the measured PL spectrum of the QWR. In order to assess the differences between a photoluminescence excitation spectrum and this optical density, we show in Fig. 3 a comparison with two PLE spectra measured at 8K and 100K. We find that the PLE spectrum measured at 100K is nearly equivalent to the optical density whereas, at 8K, it shows major deviations. We also observe a substantial decrease of the inhomogenous broadening of the optical transitions as the temperature is raised to 100K. It can be estimated precisely on the $(e_1$-$h_1)$ optical transition by performing a multi-gaussian curve fit. The standard deviation characterizing the broadening decreases from 4 to 3.2 meV. As the temperature is raised further, we find that the linewidth increases again as the homogeneous broadening originating from scattering with optical phonons starts to set in. However, an estimation based on the calculated rates for exciton-optical phonon scattering in a QWR indicates that the homogeneous contribution to the linewidth is about 0.1 meV at 100K (a total scattering linewidth of 4 meV is used) [13]. Thus, we will assume that the optical density determined at 100K is also the optical density at a lower temperature since the inhomogeneous broadening is then dominant.

The position of the optical interband transitions has been calculated in the single-particle description of quantum confinement following a model that includes valence-band mixing [14].

The set of calculated transitions has been shifted rigidly in order to make the transition ($e_1$-$h_1$) coincide with the experimental one. In this way we attempt to account for the excitonic effects, which are not included in our model. The calculated transitions are in fair agreement with the energy position of the optical transitions found in the PLE spectra. Surprisingly, we find that the energy separating the two optical transitions ($e_1$-$h_1$) and ($e_2$-$h_2$) is not the same in the two PLE spectra. The energy separation increases from 25.5 meV to 30 meV as the temperature is raised from 8K to 100K. We note that the same trend was systematically found in our temperature studies of PLE spectra for other QWR samples. We propose that interfacial disorder affects differently optical transitions that originate from the recombination of carriers confined in different subbands. This will lead to a differential Stokes shift between optical transitions in PLE spectra measured at different temperatures.The envelope wavefunction of an electron on subband n=2 has two lobes, which are displaced from the center of the crescent-shaped QWR. Large fluctuations present on the upper {311} A facets are expected to increase the disorder that is felt by an electron in the second subband. As a result the relaxation of a photoexcited electron-hole pair in such subband states will be more effectively inhibited by the stronger fluctuations at the interface. It will not contribute as efficiently to the luminescence at the set detection wavelength as an electron-hole pair excited in more extended states. We show in Fig. 4 the contour plots of the probability density of an electron in the lowest subband (n=1). A similar observation was originally made by Wang and coworkers in their analysis of PL data of a thinner crescent-shaped QWR[15].

We will now make use of the optical density to analyze the evolution of the PL spectra as the lattice temperature is varied. The spectra of Fig. 1 are fitted by using equation (2) that relates the optical density to the photoluminescence spectrum via a Boltzmann factor with the temperature as a fit parameter. The results of this analysis are presented in Fig. 5 (a) to (c). We introduced a rigid energy shift of all the optical spectra in order to set the energy zero at the maximum of the PL spectrum at each lattice temperature. The coincidence between the experimental and the calculated PL spectra is remarkable at both a lattice temperature of 80 and 140K. It remains so at 60K where the best fit indicates a temperature of 62K. We infer from this comparison that a thermal equilibrium with the lattice is indeed reached by the photoexcited electron-hole system. A quasi-equilibrium is, however, not established at a lattice temperature lower than 60K as attested to by the deviation from the thermal fit on the high energy side of the PL spectrum in Fig. 5 (c). The spectral shape of the photoluminescence at 10K and at 30K (not shown here) provides strong evidence for a non-thermal carrier distribution in steady state conditions. The deviation suggests that the carrier distribution is *hotter* in the high energy states of the QWR, which correspond to extended states. It is, however, *cooler* in the low energy states, which are strongly influenced by disorder and are thus partially localized. This picture is corroborated by simulation of exciton kinetics in a disordered potential performed by R. Zimmermann and coworkers [16]. In Fig. 6, we show the occupation factor of the electron-hole pairs distribution obtained by dividing the experimental PL spectrum by the optical density. We accounted for the temperature dependence of the band gap by introducing a rigid shift of the optical density, which is defined by the energy difference between the PLE spectra measured at 10 and 100K. On the high energy side, next to the maximum of the PL peak, the carriers are essentially distributed according to a Boltzmann factor at a temperature of 25.5 K but on the high energy side they are much hotter. The absence of a quasi-equilibrium is the result of a competition between radiative recombination and relaxation to the lower energy states. Since at this lattice temperature the average kinetic energy of the photoexcited carriers is low, the main relaxation processes correspond to carrier-scatttering with acoustical phonons via both the piezoelectric interaction and the acoustical deformation potential [17]. Relaxation by emission of acoustical phonon is expected to be strongly reduced in the energy range of the localized states. On the low energy part of the PL, the excitons are more deeply localized meaning that the extent of the excitonic wavefunction along the wire axis is spatially limited. A localization length of about 30 nm was estimated in that sample from the large radiative lifetime [18]. Because relaxation is strongly inhibited in this spectral region the occupation factor is much lower than the extrapolated Boltzmann occupation factor (see Fig. 5 (c)). Deep in the tail of the luminescence, hopping over potential barriers must occur before an exciton can reach a neighboring site at a lower energy. The reduced hopping rates at low temperature in the band-tail states of amorphous semiconductors is a well known phenomenon that is very similar to ours [19].

In order to get more insight into the relaxation dynamics of the photoexcited carriers, we performed low-density time-resolved photoluminescence experiments on the sample. From the incident photon power density, fixed at 50 W/cm$^2$, we estimate the carrier density to be about 2 x $10^4$ cm$^{-1}$ assuming an absorption coefficient of $10^4$ cm$^{-1}$ at 730 nm. In Fig. 7 we depict typical spectra at different time delays for the two lattice temperatures of 8K and 60K. In the 60 K spectra we can clearly observe a high energy tail corresponding to optical transitions $(e_1-h_4)$ , $(e_1-h_6)$ and $(e_2-h_2)$ . The identification is based on the PLE spectra of Fig. 3. Assuming that the transient emission spectra correspond to the recombination of a carrier distribution in a quasi-equilibrium we proceed to fit the spectra at different delays with the optical density determined at 100K and 100 ps after the excitation. The fitting was done over one order of magnitude on the high energy side of the optical transition ($e_1$-$h_1$). The best fits at both 8 K and 60 K are displayed respectively in Fig. 8 and 9 for two different time delays.

We find that the fits at 60 K are very good while those at 8K depart from the expected thermal spectral shape that is assumed by the fitting analysis. The results of the fits at 60 K indicate that the photoexcited electron-hole pairs are well described by a quasi-equilibrium characterized by a temperature above the lattice temperature. This quasi-equilibrium is reached very rapidly (within our temporal resolution of 10 ps) implying that carrier-carrier scattering, both for intersubband and intrasubband processes, is very effective to establish an equilibrium among the occupied subbands. We do not find any evidence of a phonon bottleneck in the relaxation dynamics at this carrier density of 2 x $10^4$ cm$^{-1}$. Furthermore, the carriers are seen to cool effectively to the lattice temperature after a time delay of about 100 ps.

The results of the fits at 8K provide clear evidence for a non-thermal distribution of the carriers at a time delay of 10 ps. The high-energy tail is found to disappear very rapidly with a time constant of 10 ps indicating that the underlying relaxation time of the electron-hole pairs to lower energy states is limited by our time resolution. At later delay times (up to about 100ps) the best fit temperature remains at about 25 K implying that the energy loss rates are very small. We believe that the presence of significant disorder affects the relaxation of the electron-hole pairs to lower energy states in these transient photoluminescence spectra at 8K and lead to non-thermal distributions of electrons and holes.

The influence of the non-Markovian electron-electron quantum kinetics in our measurements can not be inferred from our measurements as disorder-related effects are shown to play a significant role in the carrier dynamics. On the present time scale of our experiments, intersubband carrier-carrier scattering can also lead to an effective thermalization of the carriers at relatively low carrier densities as was demonstrated by Monte Carlo simulation of the carrier dynamics in QWRs [20]. Further experiments with improved time-resolution on quantum wire structures with still reduced disorder are deemed necessary to address the importance of non-Markovian quantum kinetics of electron-electron scattering in establishing a thermalized distribution.

In summary, our photoluminescence studies of a high-quality quantum wire, both for cw and pulsed excitation, establish that photoexcited electron-hole pairs do not reach a quasi-equilibrium at a lattice temperature of 8 K. A quasi-equilibrium is only found at a higher temperature exceeding 60K. Thermalization of the photoexcited electron-hole pairs is established within our time-resolution of 10 ps at 60K. These experiments have revealed the preponderant role played by disorder in the relaxation dynamics of photoexcited electron-hole pairs. They also indicated that carrier-carrier scattering processes were important in establishing a thermalized distribution in a quantum wire at a lattice temperature of 60 K and higher.

## 6 Acknowledgements

The authors wish to thank M.-A. Dupertuis for continuous stimulating discussions and for providing the band structure calculations that led to the assignement of the optical transitions. We also acknowledge F. Reinhardt for the growth of the samples. Part of this work was supported by the Swiss National Science Foundation.

**Figures**

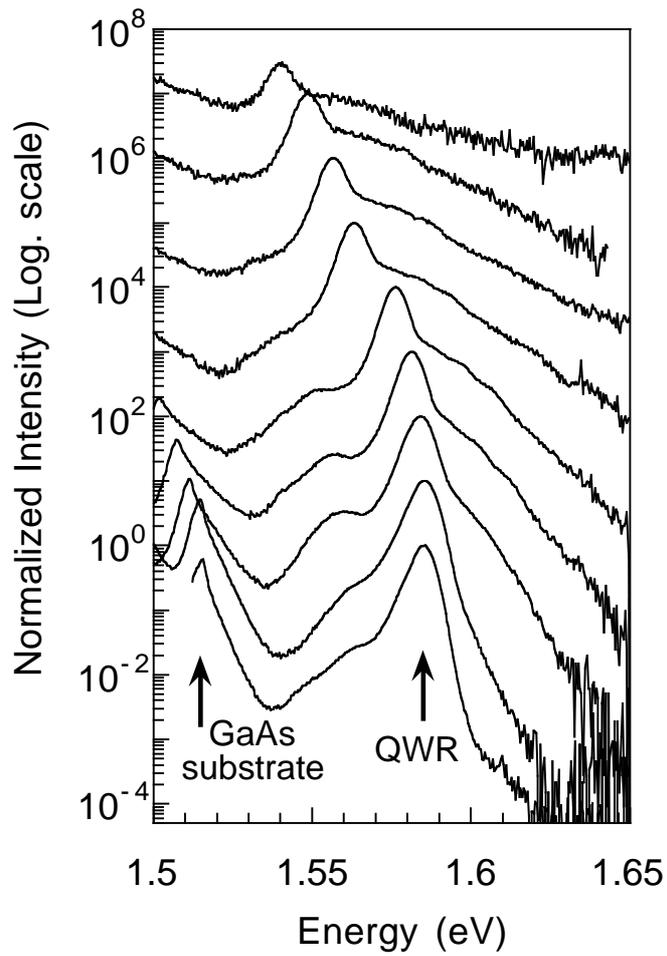

Fig. 1: Temperature dependence of the photoluminescence of a 7 nm - QWR sample. The spectra are displaced vertically for increasing temperatures (10, 30 , 60 , 80, 100, 140, 160, 180, 200 K). The QWR-peak is found at 1.585 eV in the PL-spectrum at 10K.

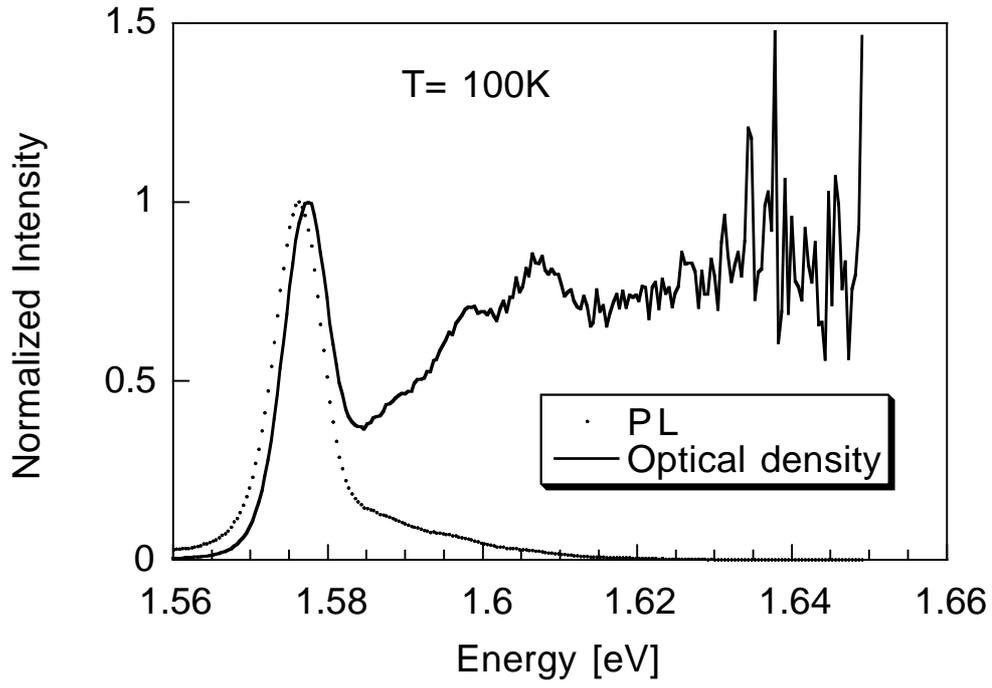

Fig. 2: PL spectrum at a lattice temperature of 100 K and corresponding optical density.

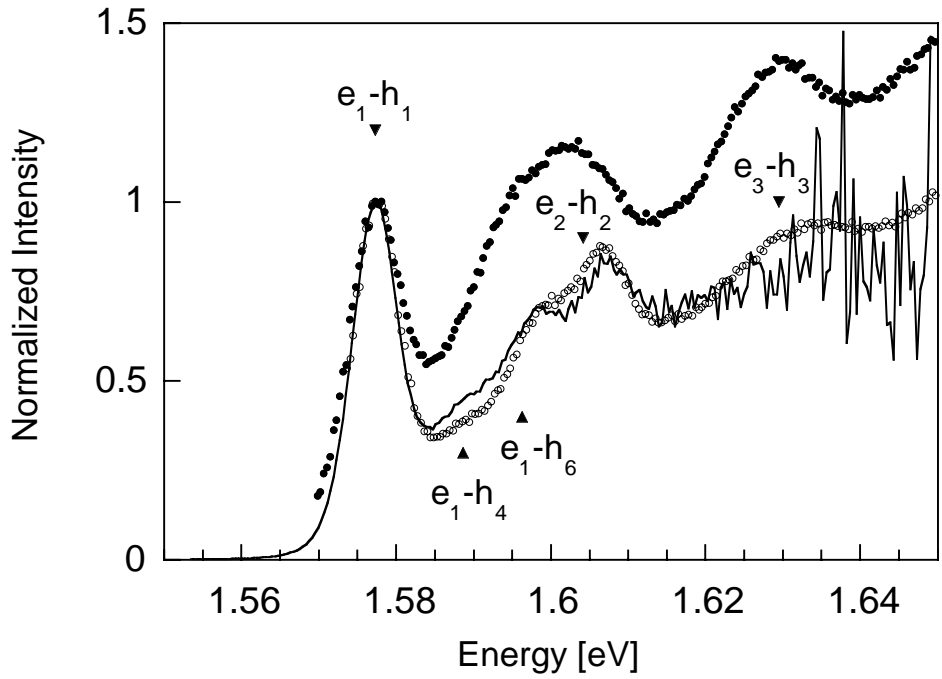

Fig. 3: Optical density spectrum of the QWR determined from the PL spectrum at 100K is displayed as a continuous line. The top and bottom PLE spectra correspond respectively to a temperature of 8K (filled circle) and 100K (empty circle). The PLE spectra are measured with circularly polarized light. The PLE spectrum at 8K has been shifted in energy such that the optical transitions $e_1$-$h_1$ coincide. The calculated positions of the different optical transitions are indicated by markers.

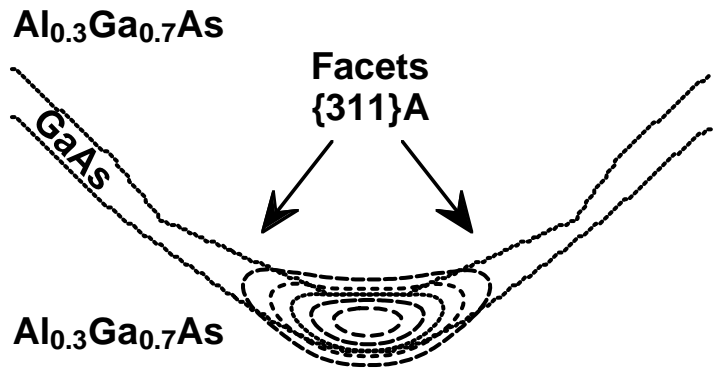

Fig. 4: Contour lines of the probability density of electron in subband (n=1). The wire cross section used for the calculation of the confinement potential is also shown. It is derived from a transmission electron micrograph of a similar (6.9 nm) QWR sample. Facets {311} A are found on the upper interface between the GaAs wire and the AlGaAs barrier surrounding the wire.

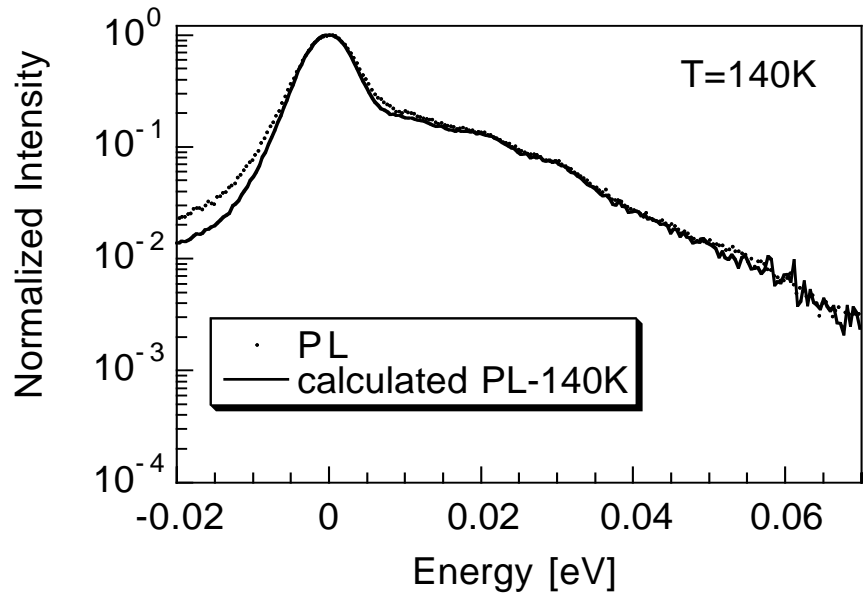

Fig. 5 (a): Comparison of the measured PL spectrum at 140K with the calculated PL spectrum. The best fit gives a temperature of 140K.

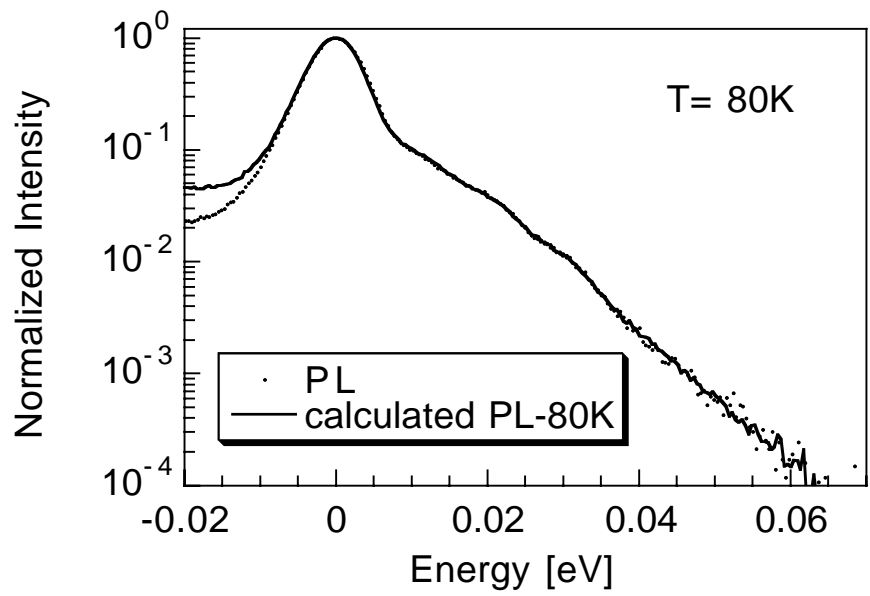

Fig. 5 (b): Comparison of the measured PL spectrum at 80K with the calculated PL spectrum.

The best fit gives a temperature of 80K.

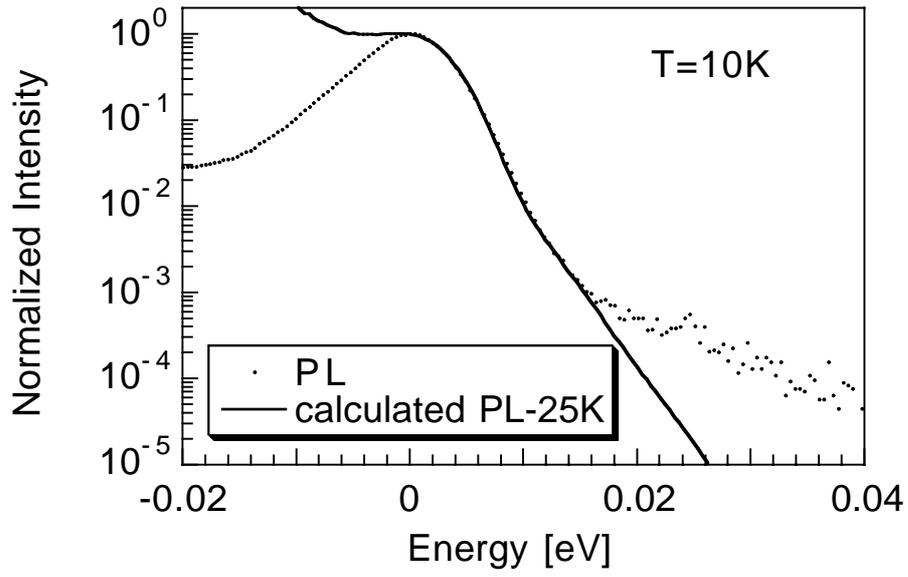

Fig. 5 (c): Comparison of the measured PL spectrum at 10K with the calculated PL spectrum. The best fit gives a temperature of 25K.

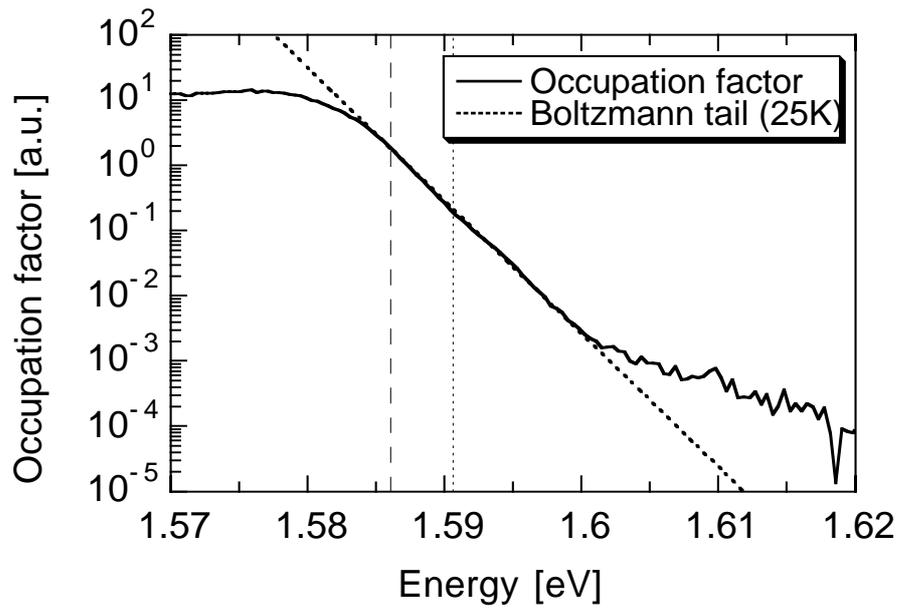

Fig. 6: Occupation factor of the photoexcited carrier distribution at a lattice temperature of 10K.

The dashed (dotted) vertical line indicates the position of the PL (PLE) peak.

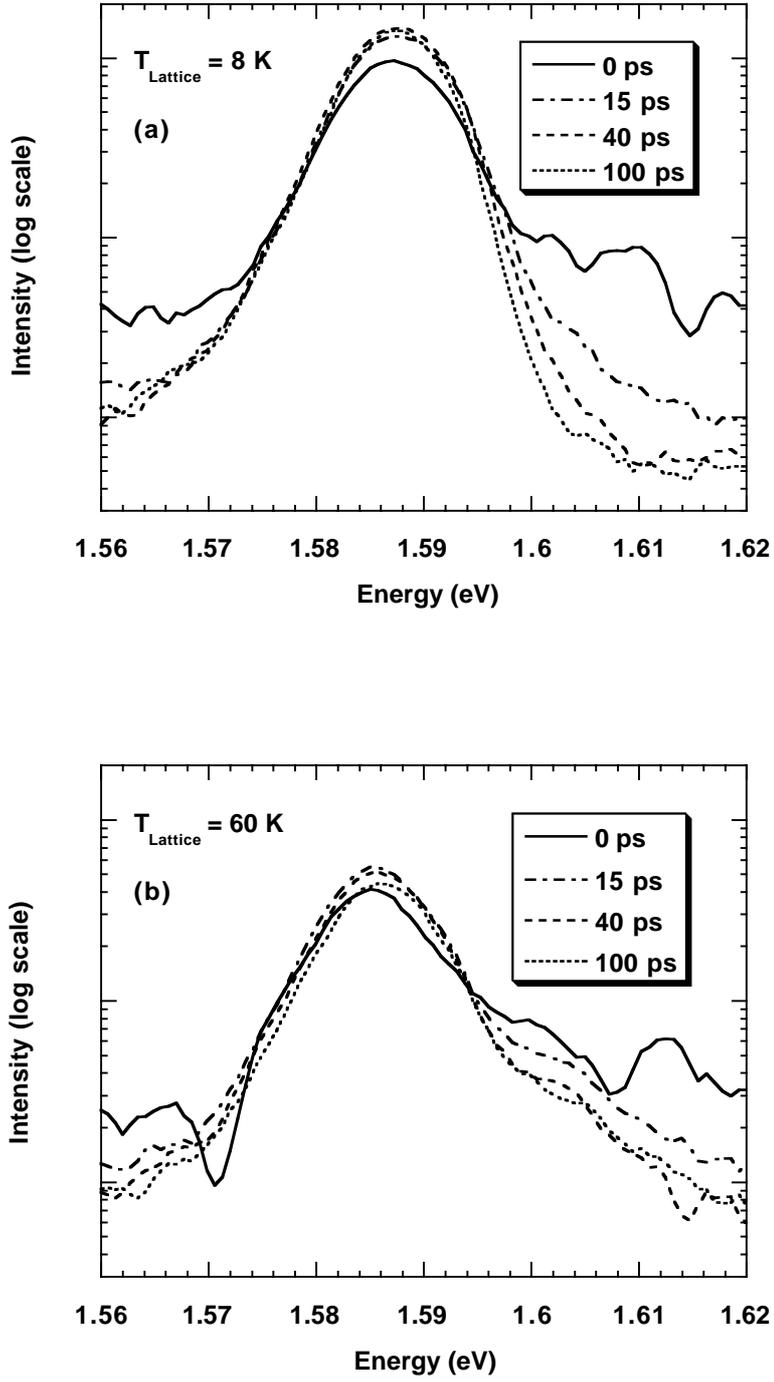

Fig. 7: Transient spectra at the two lattice temperatures of (a) 8K and (b) 60K.

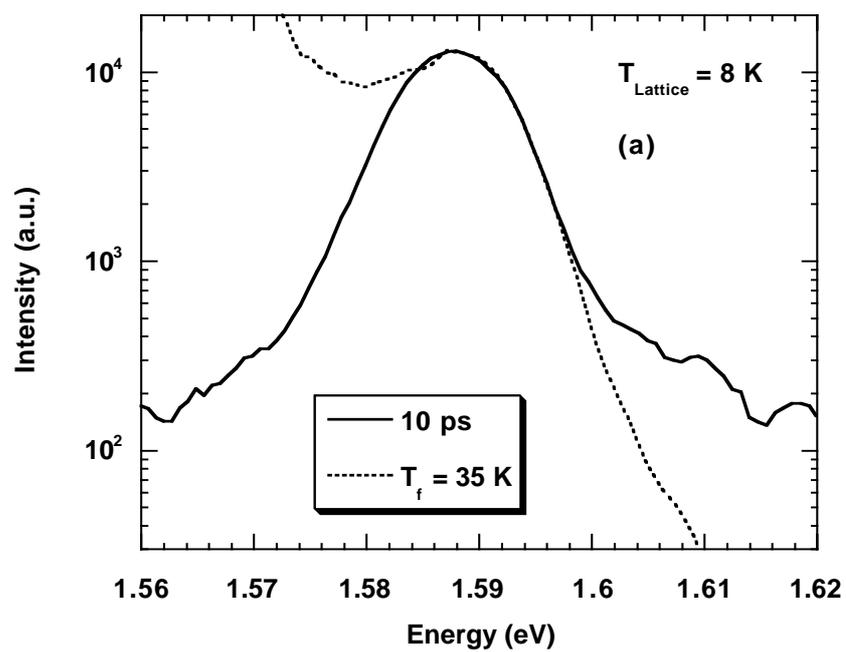

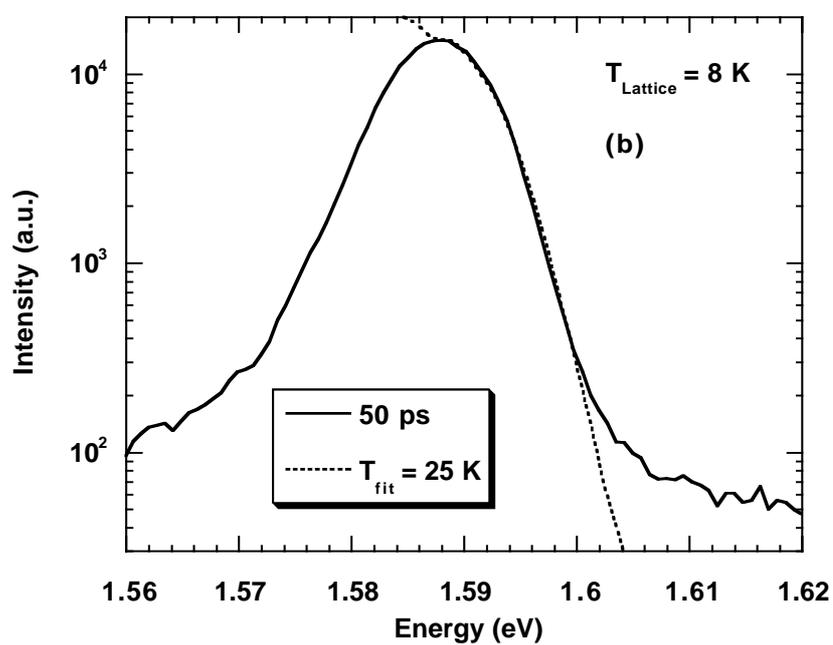

Fig. 8: Transient spectra of 8K at time delays of (a) 10 ps and (b) 50 ps together with the best

fitting curves.

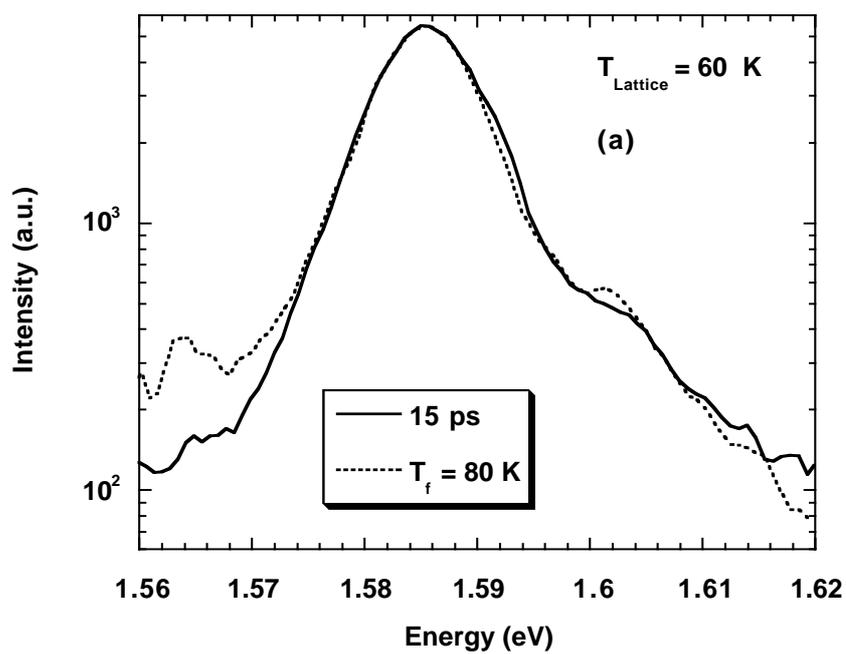

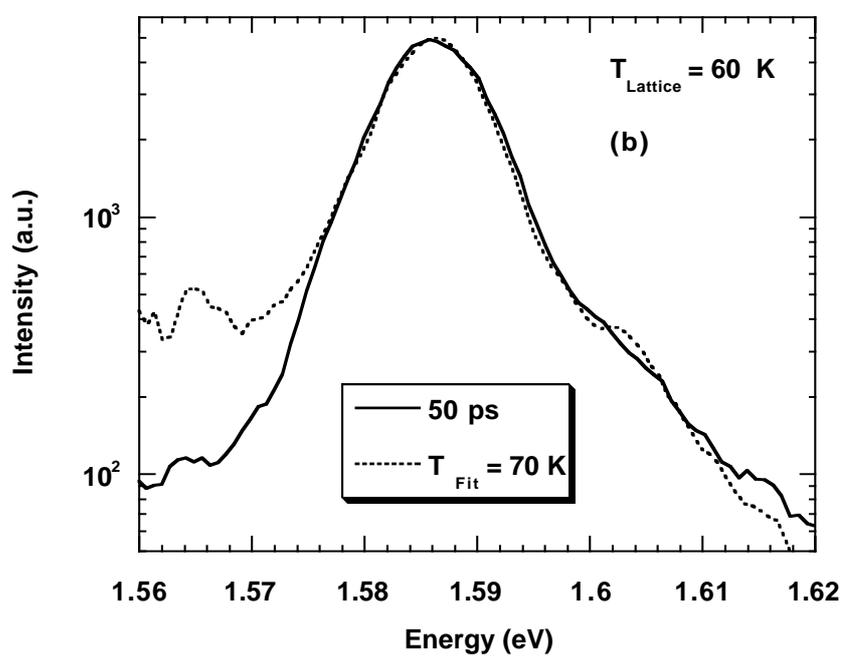

Fig. 9: Transient spectra of 60K at time delays of (a) 10 ps and (b) 50 ps together with the best

fitting curves.

**References**

1 D. S. Citrin, Phys. Rev. B 47, 3832 (1993).

2 J. Hasen et al., Nature 390, 54 (1997).

3 J. Bellessa et al., Appl. Phys. Lett. 71, 2481 (1997).

4 F. Vouilloz et al., Physica E 2, 862 (1998).

5 F. Prengel and E. Schöll, Phys. Rev. B 59, 5806 (1999).

6 The quartic term in the electron dispersion is actually sufficient to restore intrasubband electron-electron scattering processes. Because these processes only involve a pair of electrons close to the band extrema of the dispersion, they cannot lead to the thermalization of an electron distribution close to the center of the Brillouin zone (at the $\Gamma$ point).

7 G. Fasol and H. Sakaki, Phys. Rev. Lett. 70, 3643 (1993).

8 Yuri M. Sirenko, V. Mitin and P. Vasilopoulos, Phys. Rev. B 50, 4631 (1994).

9 L. Rota et al., Phys. Rev. B. 47, 1632 (1993).

10 A. Gustafsson, F. Reinhardt, G. Biasiol, and E. Kapon, Appl. Phys. Lett. **67**, 3673 (1995).

11 D. Y. Oberli, F. Vouilloz, and E. Kapon, phys. stat. sol. (a) 164, 353 (1997).

12 H. Haug and S. Schmitt-Rink, Prog. Quant. Electr. Vol. 9, 3 (1984).

13 S. Nojima, Phys. Rev. B 46, 2302 (1992).

14 F. Vouilloz et al., Phys. Rev. B 57, 12378 (1998).

15 X. Wang et al., Appl. Phys. Lett. 71, 2130 (1997).

16 R. Zimmermann, E. Runge, and F. Grosse, Proc. 23nd ICPS Berlin, Eds. M. Scheffler and R. Zimmermann (World Scientific, 1996) p. 1935.

17 M. Pugnet, J. Collet and A. Cornet, Solid State Comm. 38, 531 (1981).

18 D. Y. Oberli et al., Phys. Rev. B 59, 2910 (1999).

19 D. Monroe, Phys. Rev. Lett. 54, 146 (1985).

20 L. Rota, F. Rossi, P. Lugli, and E. Molinari, Phys. Rev. B 52, 5183 (1995).